\documentstyle[twocolumn,prl,aps,epsfig,
]{revtex}

\catcode`@=11
\def\citer{\@ifnextchar [{\@tempswatrue\@citexr}{\@tempswafalse\@citexr[]}}
 
\def\@citexr[#1]#2{\if@filesw\immediate\write\@auxout{\string\citation{#2}}\fi
  \def\@citea{}\@cite{\@for\@citeb:=#2\do
    {\@citea\def\@citea{--\penalty\@m}\@ifundefined
       {b@\@citeb}{{\bf ?}\@warning
       {Citation `\@citeb' on page \thepage \space undefined}}%
\hbox{\csname b@\@citeb\endcsname}}}{#1}}
\catcode`@=12

\def\beq{\begin{equation}}
\def\eeq{\end{equation}}

\def\beqn{\begin{eqnarray}}
\def\eeqn{\end{eqnarray}}
\relax
\def\ba{\begin{array}}
\def\ea{\end{array}}

\def\be{\begin{equation}}
\def\ee{\end{equation}}
\def\bea{\begin{eqnarray}}
\def\eea{\end{eqnarray}}

\def\to{\rightarrow}

\def\tolr{\leftrightarrow}
\def\dis{\displaystyle}
\def\f{\frac}
\def\cbar{\bar c}
\def\bbar{\bar b}
\def\tbar{\bar t}

\def\bbar{\bar{b}}
\def\tbar{\bar{t}}
\def\CL{{\cal C}_L}
\def\CR{{\cal C}_R}

\def\sq2{\sqrt{2}}
\def\gaga{\gamma\gamma}
\def\tanb{\tan\hspace*{-1mm}\beta}

\def\thisday{~hep-ph/9810367 ~and~ Phys. Rev. Lett., in press~~}
\begin{document}                                                              
\draft

\twocolumn[\hsize\textwidth\columnwidth\hsize\csname
@twocolumnfalse\endcsname
     
\title{New Method for Detecting Charged (Pseudo-)Scalars at Colliders}  
\author{{\sc Hong-Jian He} ~~and~~ {\sc C.--P.~Yuan}}
\address{\phantom{ll}}
\address{
Department of Physics and Astronomy,
Michigan State University, East Lansing, Michigan 48824, USA 
}
\date{\thisday}
\maketitle
\begin{abstract}
\hspace*{-0.35cm}
We propose a new method for detecting a charged (pseudo-)scalar 
at colliders, based upon the observation that its Yukawa coupling to 
charm and bottom quarks can be large due to a significant mixing 
of the top and charm quarks. After analyzing the typical flavor mixing 
allowed by low energy data in the topcolor and the generic two-Higgs 
doublet models, we study the physics potential of the
Tevatron, LHC, and linear colliders for probing such an $s$-channel 
charged resonance via the single-top (as well as $W^\pm h^0$) production. 
We show that studying its detection at colliders can also provide 
information on the dynamics of flavor-changing neutral current phenomena.
\linebreak
{PACS number(s): 13.85.Ni  12.60.Fr  14.65.Ha  14.80.Cp 
          \hfill [MSUHEP-80801]}
\end{abstract}
%\pacs{PACS number(s): 13.85.Ni  12.60.Fr  14.65.Ha  14.80.Cp}
\vskip1pc]
%]

%\begin{narrowtext}

%\vspace*{-0.8cm}
The large mass of the top quark ($t$)
suggests that it may play a special role in the
dynamics of the electroweak symmetry breaking (EWSB) and/or 
the flavor symmetry breaking.
The topcolor models \cite{topC,topCrev}
and the supersymmetric theories with radiative breaking \cite{SUSY0}
are two of such examples, in which 
the Higgs sector generically contains at least two 
(composite or fundamental) scalar doublets, 
and hence predicts the existence of physical charged (pseudo-)scalars
as an unambiguous signal beyond the standard model (SM).
We point out that a large flavor mixing (FM) between the 
right-handed top and charm quarks can induce a large FM Yukawa
coupling of a charged (pseudo-)scalar 
with charm ($c$) and bottom ($b$) quarks.
This is different from the usual Cabibbo-Kobayashi-Maskawa (CKM)
mixing which involves only left-handed fermions in the 
charged weak current.
Furthermore, when the neutral scalar ($\phi^0$) and the charged scalar 
($\phi^\pm$) form an SU(2) doublet, 
the weak isospin symmetry connects the 
flavor-changing neutral coupling (FCNC) $\phi^0$-$t$-$c$
to the flavor-mixing charged coupling (FMCC) $\phi^\pm$-$c$-$b$
through the CKM matrix. Hence, a direct measurement of the FMCC 
at high energy colliders can also provide information on the FCNC, 
which may give better constraint than that inferred from the low
energy kaon and bottom physics.

In this Letter, we show that with a large FMCC $\phi^\pm$-$c$-$b$,
$\phi^\pm$ can be copiously produced
via the $s$-channel partonic process $c\bbar ,\cbar b \to \phi^\pm$
at colliders, such as the Fermilab Tevatron, CERN Large Hadron Collider
(LHC) and electron/photon linear colliders (LCs).
After analyzing the production rates of $\phi^\pm$ as a function of its
mass $m_{\phi}$ and its coupling strength to 
the $c$ and $b$ quarks, we discuss the typical range of these parameters 
(allowed by the low energy data)
%including the flavor-changing neutral current constraints)
in the topcolor model (TopC) \cite{topC} and the generic 
two-Higgs doublet model (2HDM) \cite{ansatz,2HDM3}.
We show that with a significant mixing 
between the right-handed top and charm quarks,
a sizable coupling of $\phi^\pm$-$c$-$b$ can be induced from 
a top-mass-enhanced $\phi^\pm$-$t$-$b$ Yukawa coupling,  so that 
Tevatron can probe the charged top-pion mass up to $\sim$\,300--350GeV
in the TopC model, LHC can probe the mass-range of charged Higgs bosons 
up to $\sim O(1)$~TeV, and the high energy $\gaga$ LCs 
can also be sensitive to such a production mechanism.

\vspace*{0.25cm}
\noindent
\underline{\it $s$-Channel Production of Charged (Pseudo-)Scalars} 
%\vspace*{0.15cm}

With a large FM coupling of $\phi^\pm$-$c$-$b$, it is possible
to study the charged scalar or pseudo-scalar 
$\phi^\pm$ via the partonic $s$-channel production mechanism,
~$c\bbar ,\cbar b\to\phi^\pm $. 
Defining the $\bar{q}$-$q^{\prime}$-$\phi^\pm $ coupling as 
{\small $~\CL\widehat{L}+\CR\widehat{R}~$}
in which 
%${\small \Gamma} =1$~or~$\gamma_5$ and 
{\small $~\widehat{L}(\widehat R)=(1\mp\gamma_5)/2$}, 
we derive the total cross section for $\phi^+$ production
at hadron colliders as
\vspace*{-1.5mm}
\be
\ba{l}
\sigma \left(h_1h_2(c\bbar)\to \phi^+X\right)=
\dis\f{\pi}{12S}\left(|\CL|^2+|\CR|^2\right)
\times\\[1mm]
\hspace*{-3mm}
\dis\int_{\ln\hspace*{-1mm}\sqrt{\tau_0}}^{
-\ln\hspace*{-1mm}\sqrt{\tau_0}}dy
\left[f_{c/h_1}(x_1,Q^2)f_{\bbar /h_2}(x_2,Q^2)+(c\tolr\bbar )\right],
\\[-1.5mm]
\ea
\label{eq:crosstot}
\ee
where  $\sqrt{S}$ is the collider energy,
{\small $\tau_0=m_\phi^2/s ,~
x_{1,2}\hspace*{-1mm}=\hspace*{-1.2mm}\sqrt{\tau_0}
\hspace*{1mm}\dis e^{\pm y}$,~} 
and $f_{q/h}(x,Q^2)$ is the parton distribution function (PDF) with
$Q$ the factorization scale (chosen as $m_{\phi}$). 
% Here the terms suppressed by the small mass ratio
% $(m_{c,b}/m_\phi )^2$ have been ignored. 
Similarly, we can derive the
cross section formula for $~e^-e^+(\gaga )\to \phi^+ \cbar b~$
and $~\gaga\to \phi^+ \cbar b~$ at electron and photon linear colliders.
For the $e^-e^+$ process we have used the Williams-Weizsacker 
equivalent photon approximation.      %~\cite{EPA}.
The present analysis for the signal event is confined to the tree level,
and the CTEQ4L PDFs are used for hadron collisions. 
The complete next-to-leading order (NLO) QCD correction 
(including the $b(c)$-gluon fusions) 
will improve the numerical results but will not change our main 
conclusion~\cite{QCD}.
   
To illustrate the $\phi^\pm$ production rates at various colliders, we
consider its Yukawa couplings to be the typical values of TopC 
models [cf. eqs.~(\ref{eq:KURtc}-\ref{eq:Ltoppi}) below]:
{\small $\CL^{tb}=\CL^{cb}=0$} 
and 
{\small $\CR^{tb}=y_{t0}\tanb ,~\CR^{cb}\simeq \CR^{tb}\times 0.2$,} 
with  $y_{t0}=\sqrt{2}m_t/v$, $\tanb \simeq 3$ and $v\simeq 246$\,GeV,
which serves as a benchmark of our general analysis. 
In Fig.\,1 we plot the $s$-channel resonance cross section versus
the mass of $\phi^\pm$ at hadron and electron/photon colliders.
%(A factor 2 is added to eq.~(1) for including the rate of $\phi^-$.) 
For $m_\phi =200 \, [1000]$~GeV at the 1.8 and 2~TeV Tevatron
[14~TeV LHC], the total 
cross sections are 2.7 and 4.0 [0.55]~pb, respectively.
The production rates, calculated from a complete gauge invariant
set of $c$-$b$ fusion diagrams, 
are also large at the $\gamma$-$\gamma$ LCs.
[Here, we do not include the production 
of $\phi^\pm$-pair with one scalar
decaying into $bc$, whose
cross section is large for $m_\phi \ll \sqrt{S}/2$.]
It is trivial to rescale our results 
to other values of ${\cal C}_{L,R}$.
For instance, the typical couplings of the generic 2HDM 
[cf. eqs.~(\ref{eq:2HDM3}-\ref{eq:ansatz})] are
{\small $\CL^{tb}=\CL^{cb}\simeq 0$} and 
{\small $\CR^{tb}=\xi^U_{tt},
         ~\CR^{cb}\simeq \xi^U_{tc}\times 9\%$.}\,
Taking the sample value {\small $\xi^U_{tc}\sim 1.5$}, 
we find that a typical prediction of this 2HDM can
be obtained from rescaling the solid curves in
Fig.\,1 by a factor of $1/19$.
\begin{figure}
\vspace*{-1.1cm}
\hspace*{-1cm}
%\vspace*{6.5cm}
\epsfxsize=10cm\epsfysize=9.5cm
\epsfbox{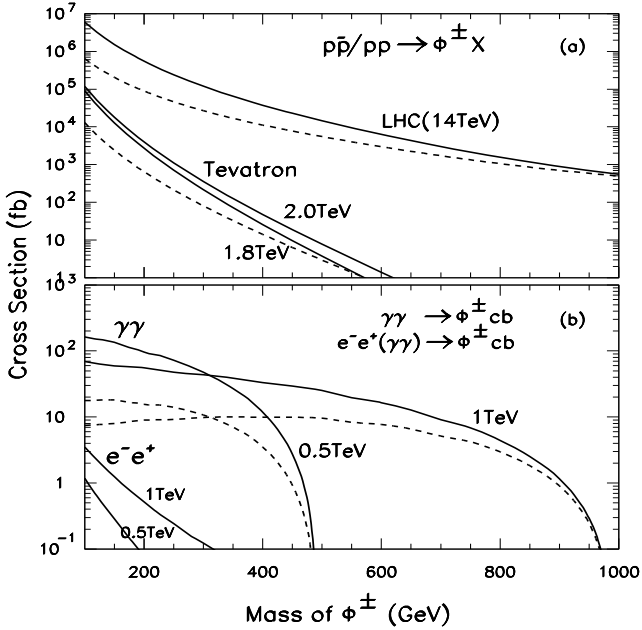}\\[-0.7cm]
\caption[fig:fig1]{
The $s$-channel $\phi^\pm$-production 
%at (a) hadron and (b) electron/photon colliders 
with the benchmark parameter choice of TopC models. 
As a reference, dashed curves show the results  
at Tevatron\,(2\,TeV), LHC and $\gaga$ LCs, 
with top-pion Yukawa couplings satisfying the roughly estimated 
$3\sigma$ $R_b$-bound\,\cite{Rb}
(reanalyzed with new $R_b^{\rm exp}$ data).
} 
\label{fig:sigmatot}
\end{figure}
\begin{figure}
\vspace*{-1.6cm}
\hspace*{-1cm}
%\vspace*{6.5cm}
\epsfxsize=10cm\epsfysize=8.7cm
\epsfbox{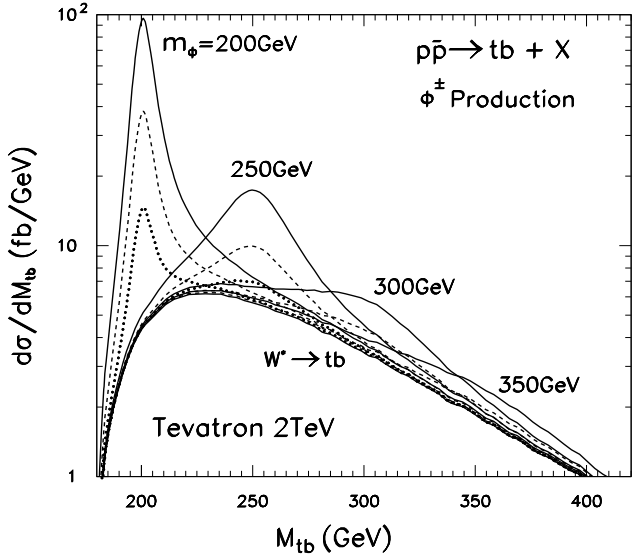}\\[-0.7cm]
\caption[fig:fig2]{$tb$ invariant-mass distribution  
of charged top-pion production, 
with its Yukawa coupling constrained by 
the estimated $3\sigma$ $R_b$-bound. The (solid,\,dashed,\,dotted)
curves are plotted for 
${\cal C}_R^{cb}={\cal C}_R^{tb}\times (0.33,0.2,0.11)$, 
corresponding to 3 typical values of $t_R$-$c_R$ mixing 
parameter $K_{UR}^{tc}$ in eq.\,(6).
%Including the top-polarization
%measurement can reduce the $W^\ast$ background substantially.
} 
\label{fig:Mtb}
\end{figure}
%\noindent
\vspace*{-3mm}
For $m_\phi > 190$~GeV, the dominant decay mode of 
$\phi^\pm$ can be the $t{\bar b}$ pair, 
which is the case for TopC-type of models (cf. Fig.~3). Therefore,
$\phi^\pm$ may be detected via single-top production.
The $\phi^\pm$ production rate at the Tevatron drops very
fast for $m_\phi \geq 300$~GeV and becomes comparable with the SM
$s$-channel single-top rate, via
$q\bar{q}^{\prime}\to W^{\ast}\to t\bbar ,\tbar b$. 
[The NLO $W^\ast$ rate is about 0.70, 0.86 and 11.0~pb at the 
1.8, 2~TeV Tevatron and the 14~TeV LHC, respectively.
Although the $t$-channel single-top rate
(via $Wg$ fusion) is larger~\cite{Tim-CP}, its event topology is 
different from the $s$-channel single-top event.  Therefore, we only
compare the $\phi^\pm$ signal rate with the $W^\ast$ rate.]
Thus, an analysis on the distribution of the $t$-$b$ 
(or $W$-$b$-$\bbar$) invariant mass
will ensure the identification of $\phi^\pm$ due to 
its resonance peak (cf.~Fig.~2). 
The Tevatron Run~I data (at 1.8~TeV) may already put 
important bounds on some parameter space 
of $m_\phi$ and ${\cal C}_{L,R}$.

\vspace*{0.25cm}
\noindent
\underline{\it Flavor-Mixing and Top-pion $\pi_t^\pm$ Production in TopC}
\vspace*{0.15cm}

The topcolor scenario \cite{topC,top-seesaw} is attractive because it
explains the large top quark mass and provides possible dynamics of the
EWSB. Such type of models generally predict 
light composite (pseudo-)scalars with large
Yukawa coupling to the third family. This induces distinct
new FM phemonena which can be tested at both low and high energies.
In the typical topcolor-I class of models \cite{Hill2}, three light
pseudo-scalars, called top-pions, are predicted with masses around 
of $O(150\hspace*{-0.4mm}-\hspace*{-0.4mm}300)$GeV. 
The up(down)-type quark mass matrices 
$M_U$($M_D$) exhibit an approximate triangular texture due to the
generic topcolor breaking pattern \cite{Hill2}. 
For generality, we can write %$M_U$ as
\\[-4mm]
\be
M_U~=~\left( 
\ba{ccc} m_{11}   & m_{12}   & \delta_1\\
         m_{21}   & m_{22}   & \delta_2\\
         \delta_3 & \delta_4 & m_t^{\prime}
\ea
\right)
\label{eq:MU}
\ee
\\[-3mm]
where the small non-diagonal pieces 
$\delta_{1,2}=O(\epsilon^4)v$, $\delta_{3,4}=O(\epsilon)v$
and thus the 33 element $m_t^{\prime}$ is very close to the physical 
top mass $m_t$.
Here, $\epsilon = O(\langle\Phi\rangle /\Lambda_0)$,
with $\Lambda_0$ being the breaking scale of a larger group 
down to the topcolor group, and the vacuum expectation
value $\langle\Phi\rangle$ being the topcolor
breaking scale. So, we expect $\epsilon < 1$.
A proper rotation of the quarks from weak eigenstates into mass 
eigenstates will 
diagonalize $M_U$ and $M_D$, so that  
{\small $~K_{UL}^\dag M_U K_{UR} = M_U^{\rm dia}~$} and 
{\small $~K_{DL}^\dag M_D K_{DR} = M_D^{\rm dia}$},~ 
from which the CKM matrix can be derived as 
{\small $~V=K_{UL}^\dag K_{DL}~$}.
We show that it is possible to construct a realistic
but simple pattern of the left-handed rotation matrices $K_{UL}$ and
$K_{DL}$ such that 
(i) the Wolfenstein-parametrization\cite{Wolf} of CKM
is reproduced, 
(ii) Cabibbio-mixing is mainly generated from $K_{DL}$
while the mixings between the 
3rd and 2nd families are mainly from $K_{UL}$,
(iii) all dangerous contributions to low energy data (such as the 
{\small $K\hspace*{-1mm}-\hspace*{-1mm}\overline{K}$, 
$D\hspace*{-1mm}-\hspace*{-1mm}\overline{D}$} and
{\small $B\hspace*{-1mm}-\hspace*{-1mm}\overline{B}$} 
mixings and the $b\to s\gamma$ rate) can be evaded.
In fact, the point (ii) may help to explain the successful empirical 
relations {\small $~\lambda \approx \sqrt{m_d/m_s}~$} and
{\small $~V_{cb}\approx \sqrt{m_c/m_t}~$}, 
where $\lambda$ is the Wolfenstein-parameter. 
By introducing the unitary matrices
$K_{UL}$ and $K_{DL}$,
%\\[-5.5mm]
\vspace*{-1.5mm}
\be
\ba{l}
K_{UL}\hspace*{-1.2mm}=\hspace*{-1.2mm}
       \left(\ba{ccc} 1 & 0        & 0  \\
                      0 & c        & -sx\\
                      0 & sx^\ast & c
             \ea
      \right)\hspace*{-1.1mm},~~ 
K_{DL}\hspace*{-1.2mm}=\hspace*{-1.2mm}
      \left(\ba{ccc} c_\phi & s_\phi c_\theta & s_\phi s_\theta y\\
                     -s_\phi & c_\phi c_\theta & c_\phi s_\theta y\\
                      0      & -y^\ast s_\theta & c_\theta
             \ea
      \right)\hspace*{-1.1mm}, \\[-2mm]
\ea
\label{eq:KL}
\ee

\vspace*{-3.5mm}
\noindent
with $|x|=|y|=1$ and $s_j^2+c_j^2=1$, 
we find the solution to reproduce 
Wolfenstein-parametrization up to {\small $O(\lambda^3)$}:\\[-5mm]
%\begin{flushleft}
\be\ba{l}
c,c_\theta =1+O(\lambda^4),~~ c_\phi =1-\lambda^2/2,~~
s_\phi =\lambda +O(\lambda^3),%\nonumber
\\[0.15cm]
{\small s=\lambda^2A[(1-\rho )^2+\eta^2]^{1/2},~~ 
s_\theta =\lambda^2A[\rho^2+\eta^2]^{1/2},}\\[0.15cm]
{\small x =(1\hspace*{-1mm}-\hspace*{-1mm}\rho +i\eta )/
[(1\hspace*{-1mm}-\hspace*{-1mm}\rho )^2+\eta^2]^{1/2},~ 
y=(\rho\hspace*{-1mm}-\hspace*{-1mm}i\eta)/
[\rho^2\hspace*{-1mm}+\hspace*{-1mm}\eta^2]^{1/2},}\\[-1.5mm]
\ea
\label{eq:KLX}
\ee
%\end{flushleft}
\\[-0.8cm]
where 
{\small $~\lambda\simeq 0.22,~ A\simeq 0.82$}, and 
{\small $\sqrt{\rho^2+\eta^2}\simeq 0.43$.~}
Given the matrices {\small $M_U$, $K_{UL}$} and the known 
${\small M_U^{\rm dia}=}{\rm diag}(m_u,m_c,m_t)$,
the right-handed rotation matrix $K_{UR}$ is constrained,
and the matrix elements \\[-4mm]
\be
\ba{l}
K_{UR}^{tt}\simeq \dis\f{m_t^\prime}{m_t}~,~~~
K_{UR}^{tc}\leq \sqrt{1-{K_{UR}^{tt}}^2}~.\\[-2mm]
\ea
\label{eq:KUR}
\ee
For the reasonable values of $\delta m_t=m_t-m_t^\prime =O(1-10$~GeV$)$,
(\ref{eq:KUR}) gives 
\be
K_{UR}^{tt}=0.99\hspace*{-0.7mm}-\hspace*{-0.7mm}0.94~,~~~
K_{UR}^{tc}\leq 0.11\hspace*{-0.7mm}-\hspace*{-0.7mm}0.33~,
\label{eq:KURtc}
\ee
which shows that the $t_R$-$c_R$ transition 
can be naturally around $10$-$30\%$. 
[It also requires $\delta_{1,2}(=O(\epsilon^4)v)=O({\rm GeV})$, 
which suggests $\epsilon = O(0.2-0.4)$.]
Since the mass hierarchy in the down-quark sector is much smaller
than that in $M_U$, the mass pattern of $M_D$ is taken to be less
restrictive.
It is easy to check that the above $K_{UL,DL}$ and $K_{UR}$ 
satisfy the requirement of the point (iii), 
in contrast to 
the naive {\small $\sqrt{\rm CKM}$}-ansatz ~\cite{Hill2}.
(The $b\to s\gamma$ rate also has a contribution 
{\small $C_7'(M_W)$} depending on {\small $K_{DR}^{bs}$} \cite{Hill2}. 
Since the pattern of $M_D$ is less certain in this model, we take 
{\small $K_{DR}^{bs}$} more or less free, {\it e.g.,} a simple
{\small $\sqrt{\rm CKM}$}-ansatz for $K_{DR}$ can already 
accommodate ${\rm BR}[b\to s\gamma ]_{\rm exp}$ data~\cite{Hill2}.)

The relevant FM vertices including the large $t_R$-$c_R$ transition 
for the top-pions can be written as 
\be
\ba{l}
\dis\f{m_t\tanb}{v}\hspace*{-1.1mm}\left[
iK_{UR}^{tt}{K_{UL}^{tt}}^{\hspace*{-1.3mm}\ast}\overline{t_L}t_R\pi_t^0
\hspace*{-1mm}+\hspace*{-1.2mm}\sq2
{K_{UR}^{tt}}^{\hspace*{-1.3mm}\ast}K_{DL}^{bb}\overline{t_R}b_L\pi_t^+ 
+ \right.
\\[2.3mm]
~~~~\left.
iK_{UR}^{tc}{K_{UL}^{tt}}^{\hspace*{-1.3mm}\ast}\overline{t_L}c_R\pi_t^0
\hspace*{-1mm}+\hspace*{-1.2mm}\sq2
{K_{UR}^{tc}}^{\hspace*{-1.3mm}\ast}K_{DL}^{bb}\overline{c_R}b_L\pi_t^+ 
\hspace*{-1mm}+\hspace*{-1mm}{\rm h.c.}       \right],
\ea
\label{eq:Ltoppi}
\ee
where {\small $\tanb = \sqrt{(v/v_t)^2-1}$} and
$v_t\simeq O(60-100)$~GeV is the top-pion decay constant.
An important feature is that the charged top-pion 
$\pi_t^\pm$ mainly couples to the right-handed top 
($t_R$) or charm ($c_R$) but not
the left-handed top ($t_L$) or charm ($c_L$), 
in contrast to the standard $W$-$t$-$b$
coupling which involves only $t_L$. Note that $\pi_t^\pm$
also has a topcolor-instanton induced coupling with $t_L$, of the
strength $\sq2 m_b^{\star}/v_t$ \cite{topC}, which is much suppressed 
by $m_b^{\star}\leq m_b \ll v_t$. The tiny left-handed rotation element
{\small $|K_{UL}^{tc}|=s\approx 2-4\%$} 
[cf. (\ref{eq:KL})-(\ref{eq:KLX})] 
further makes the $c_L$-$b_R$ coupling to $\pi_t^\pm$ negligible.             
Hence, the produced top quark from $\phi^\pm$-$t$-$b$ interaction is close
to hundred percent right-handed polarized, and measuring
the top polarization in the single-top event provides further
identification of the signal.  Eq.~(\ref{eq:Ltoppi}) suggests 
that the neutral top-pion $\pi_t^0$ can be
produced in association with the single-top via charm-gluon fusion,
i.e., $cg,\cbar g\to t\pi_t^0,\tbar\pi_t^0$.
However, due to the limited Tevatron energy, the $t$-$\pi_t^0$ production 
is only feasible at the LHC. 
This is similar to a study~\cite{Yuan} for the 2HDM, but a
much larger signal rate for $\pi_t^0$ is expected
because of the enhanced $\pi_t^0$ Yukawa coupling. 
Typically, for $m_{\pi_t^\pm}>m_t+m_b$,
the main decay channels of $\pi_t^\pm$ are $tb$ and $cb$.
The total decay width and the 
branching ratios (BRs) of $\pi_t^\pm$ are shown in
Fig.~3, separately, in which we have assumed that the 
$tb$ and $cb$ pairs are the only two available decay channels 
of $\pi_t^\pm$ up to 1~TeV, though the top-pion is not expected
to be very heavy.
[Note that the mass splitting $m_{\pi^\pm_t}-m_{\pi_t^0}$ may be 
larger than {\small $M_W$} in certain parameter region 
so that $\pi_t^\pm\to {\small W}\pi_t^0$
channel can become important as well. 
This will make the pattern of BRs for
$\pi_t^\pm$-decay similar to that of 
$H^\pm$-decay in the 2HDM (cf. Fig.~3b).]
From Figs.~1 and 3, we can estimate the single-top event 
rates at various colliders.
For the 1.8 and 2.0~TeV Tevatron  [14~TeV LHC] with 
$0.1$ and $2~[100]~$fb$^{-1}$ luminosity and $m_\phi =200~[500]$~GeV,
we find the numbers of single $t$ and $\tbar$ events to be
$153$ and $4.5\times 10^3~[1.4\times 10^6]$, 
while for the 0.5~[1.0]~TeV 
photon-photon ($\gaga$) LC with a 50~[500]~fb$^{-1}$ luminosity
the rate becomes $2.9\times 10^3$~[$1.2\times 10^4$] for 
$m_\phi=200~[500]$~GeV.
(If top decays semileptonically, a branching ratio of $21\%$
for $t\to bW(\to \ell\nu_\ell )$, with $\ell=e \, {\rm or} \, \mu$,
should be included.)
\begin{figure}
\vspace*{-8mm}
\hspace*{-1.45cm}
\epsfxsize=10cm\epsfysize=6cm
\epsfbox{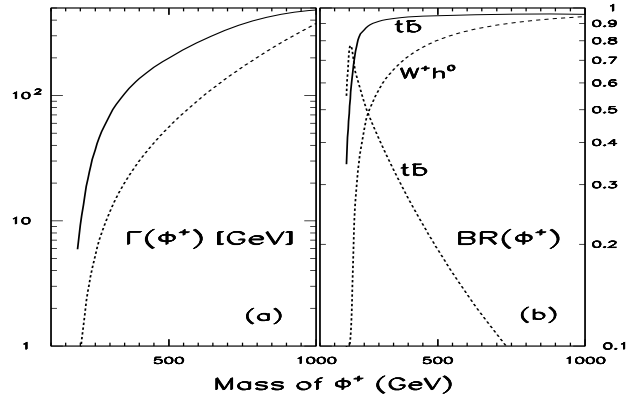} %\\[-5mm]
\vspace*{-0.5mm}
\caption[fig:fig3]{
Total decay widths (a)
and branching ratios (b)
of $\pi_t^\pm$ in the TopC model (solid), 
and of $H^\pm$ in the 2HDM (dashed) for the typical
parameter choice in the text.
} 
\label{fig:TopCWDBR}
\end{figure}

%\vspace*{0.25cm}
\noindent
\underline{\it Flavor-Mixing and Charged Higgs Production in 2HDM}
\vspace*{0.15cm}

The 2HDM is the simplest extension of the SM
and its supersymmetrization results 
in the popular minimal supersymmetric
SM (MSSM). In contrast to the MSSM,
the general 2HDM has potentially dangerous tree-level 
flavor-changing neutral currents.  
Under the reasonable ansatz for the structure 
of the Higgs Yukawa couplings
[cf.~(\ref{eq:ansatz})], the pattern for the FM Yukawa
couplings may be naturally generated from the quark-mass hierarchy, 
which sufficiently suppresses the FCNC for the light generations 
while predicting significant mixings between the charm and 
top quarks. The generic 2HDM considered here is called ``type-III''
\cite{ansatz,2HDM3}, 
which does not make use of the {\it ad hoc} discrete
symmetry~\cite{G-W}. 
The FM Yukawa couplings can be conveniently formulated under a
proper basis of Higgs doublets such that 
{\small $\langle\Phi_1\rangle =(0,v/\sq2 )^T$} and
{\small $\langle\Phi_2\rangle =(0,0)^T$.}
Thus, the diagonalization of the fermion mass matrix also 
diagonalizes the Yukawa couplings of $\Phi_1$,  
and all the FM couplings
are generated by $\Phi_2$. The Yukawa interaction of the quark sector
can be written as
\be
\ba{ll}
-{\cal L}_Y^q = 
& \f{\sq2}{v}\left[M_{ij}^U\overline{Q_{iL}}\widetilde{\Phi}_1u_{jR}+
                   M_{ij}^D\overline{Q_{iL}}\Phi_1d_{jR}\right]+\\[2mm]
&        ~~~~\left[Y_{ij}^U\overline{Q_{iL}}\widetilde{\Phi}_2u_{jR}+
                   Y_{ij}^D\overline{Q_{iL}}\Phi_2d_{jR}\right]+{\rm h.c.}
\ea
\label{eq:2HDM3}
\ee
Here the Higgs boson states are 
{\small $(H_1^0,H_2^0,A^0,H^\pm )$} and the 
$CP$-even neutral states
{\small $(H_1^0,H_2^0)$} rotate into the mass eigenstates
$(h^0,H^0)$, characterized by the mixing angle $\alpha$.
The $t$-$b$-$H^\pm$ and $c$-$b$-$H^\pm$ interactions are:
\be 
\ba{l}
H^+[\overline{t_R}~(\widehat{Y}_U^\dag V)_{tb}~b_L
        -\overline{t_L}~(V\widehat{Y}_D)_{tb}~b_R
   ]~+
\\[1.5mm]
H^+[\overline{c_R}~(\widehat{Y}_U^\dag V)_{cb}~b_L
        -\overline{c_L}~(V\widehat{Y}_D)_{cb}~b_R ~
   ]+{\rm h.c.}
\ea
\label{eq:Hcb}
\ee
where {\small 
$\widehat{Y}^{U(D)}=K_{U(D)L}^{\hspace*{1mm}\dag}Y^{U(D)}K_{U(D)R}$ }
and $V$ is CKM matrix.    The couplings 
{\small $\widehat{Y}_{ij}^{U,D}$} 
thus contain all new FM effects and exhibit a natural hierarchy
under the ansatz~\cite{ansatz,2HDM3}
\be
\dis \widehat{Y}_{ij}^{U,D}=\xi_{ij}^{U,D}
{\sqrt{m_im_j}}/
{<\hspace*{-1.2mm}\Phi_1\hspace*{-1.2mm}>}
\label{eq:ansatz}
\ee
with {\small $~\xi_{ij}^{U,D}\sim O(1)$}.
(\ref{eq:ansatz}) shows that the $t$-$c$ 
or $c$-$t$ transition gives the 
largest FM coupling while the FMs without 
involving the top quark are highly
suppressed by the light quark masses. 
Such a suppression is shown to 
persist at high energy scales according 
to a recent renormalization group
analysis~\cite{RG-2HDM3}.  
From (\ref{eq:Hcb}) and (\ref{eq:ansatz}), we deduce
{\small $~(\widehat{Y}_U^\dag V)_{cb}\simeq\widehat{Y}^{U\ast}_{tc}V_{tb}
+\widehat{Y}^{U\ast}_{cc}V_{cb}
\simeq \widehat{Y}_{tc}^{U\ast}~$}
and {\small $~(V\widehat{Y}_D)_{cb}\simeq \widehat{Y}_{sb}^D\ll
\widehat{Y}_{tc}^{U\ast}.$} 
Therefore, the dominant FM vertex $c$-$b$-$H^\pm$ involves
$c_R b_L$ but not $c_Lb_R$.
%It was found \cite{2HDM3,Laura} that low energy data allow 
It was found \cite{2HDM3} that low energy data allow 
$~ 
\xi^U_{tc},\xi^U_{ct}\sim O(1)~
$
and require  
$~
m_{h,H} \leq m_{\pm} \leq m_A ~{\rm or}~
m_{A}   \leq m_{\pm} \leq m_{h,H},~
~$
where $m_{\pm}$ is the mass of $H^\pm$ and $m_{h,H,A}$ the 
masses of $(h^0,H^0,A^0)$.
(The Higgs mixing angle $\alpha$ is not constrained.) 
For $~m_\pm > m_t+m_b$,~  $H^\pm$ can decay into the
$tb$ and $cb$ pairs. If $m_\pm >M_W+m_h$ or $m_\pm >M_W+m_A$, 
then additional decay channels, such as $H^\pm \to W^\pm h^0,~
W^\pm A^0$, should also be considered.
The total decay width and the branching ratios of $H^\pm$ for a
typical parameter set of $(\xi^U_{tt},\xi^U_{tc})=(1,1.5)$, 
$m_h=120\,$GeV, $m_A\geq m_\pm$ and $\alpha=0$
are shown in Fig.~3, separately. 
[We have verified that the choice of $\xi^U_{tt}=1$ is 
consistent with the current
$3\sigma$ $R_b$-bound for $m_\pm \geq 120$\,GeV.
Choosing a larger value of $\xi^U_{tt}>1$ will 
simutaneously increase (reduce) the BR of $tb$ ($Wh^0$) mode.]
At the 14\,TeV LHC with a 100\,fb$^{-1}$
luminosity and for $m_\phi =300\,[800]$\,GeV, about 
$1.3\times 10^5~[384]$ single-top  
and  $1.8\times 10^5~[4.1\times 10^3]$ $Wh^0$ events
will be produced, while the 0.5\,[1.0]\,TeV $\gaga$ LC
with a 50\,[500]\,fb$^{-1}$ luminosity can 
produce about 52~[131] single-top and 73~[545] $Wh^0$ events
for $m_\phi =300\,[500]$\,GeV.    From Fig.\,3b,
we note that the $t b$ and the $W^\pm h^0$ decay modes
are complementary in low and high mass ranges of $H^\pm$, 
though the details may depend on 
the values of $(m_h,m_A)$ and $\alpha$. 
Therefore, in this model it is possible to detect $H^\pm$
in either the single-top or  
the $Wh^0(\to b\bbar, \tau \tau )$ event.

%\vspace*{0.3cm}
%\noindent
%\underline{\it Conclusions}
%\vspace*{0.25cm}

In summary, the $s$-channel production mechanism proposed in this
work provides a unique probe of the charged (pseudo-)scalars at the
hadron and electron/photon colliders. 
We focus on probing the $s$-channel charged scalar 
or pseudo-scalar via single-top production 
(as well as $W^{\pm}h^0$), 
which is generic for the
TopC model and the type-III 2HDM.
For other models (such as supersymmetric theories),  
the leptonic decay channel ({\it e.g.}, $\tau\nu_\tau$ mode)
may become significant as well, and thus should   
be included. At the upgraded Tevatron, we 
show that under the typical flavor-mixing pattern of the 
TopC models, the charged top-pion mass can be explored up to 
$\sim$300-350\,GeV;  while the LHC can probe the charged Higgs
mass up to $\sim O(1)$\,TeV for the 2HDM. The linear colliders,
especially the $\gaga$ collider, 
will be effective for this purpose even
at its early phase with $\sqrt{S}=500$~GeV. 
For the MSSM, supersymmetry forbids 
tree-level FCNCs, and the 
small FMCCs are described by the usual CKM mixings. 
However, a large FMCC may be induced 
at the loop level, depending on the 
soft-breaking parameters of the model.
Work along this line is in progress.

%\newpage
\smallskip 
We thank C.~Balazs, J.L.~Diaz-Cruz, L.~Reina and T.~Tait for discussions.
CPY thanks S.~Chivukula and H.~Georgi for asking an 
interesting question on top-pion decay at Aspen, 
and HJH is grateful to M.E.~Peskin for stimulating discussions
at Keystone. 
This work is supported by the U.S.~NSF under grant PHY-9802564.\\[-0.9cm]

%\end{narrowtext}

\end{document}